# Study of a Novel Capacitive Pressure Sensor Using Spiral Comb Electrodes

Wenjie Chen[1], Qi Yang[1], Qi Liu[1], Yiqun Zhang[1], Liang He[1], Yuanlin Xia[1], Zhuqing Wang[1,2]

Yubo Huang, Jianfeng Chen[3,*], Cao Xia[1,*]

[1] School of Mechanical Engineering, Sichuan University, Chengdu 610065, China.
[2] Med+X Center for Manufacturing, West China Hospital, Sichuan University, Chengdu 610041, China.
[3] Advanced Institute for Materials Research (WPI-AIMR), Tohoku University, Sendai 980-8577, Japan

*Correspondence: xiacao_30@scu.edu.cn; ericonium@tohoku.ac.jp

**Abstract**

For traditional capacitive pressure sensors, high nonlinearity and poor sensitivity greatly limited their sensing applications. Hence, an innovative design of capacitors based on spiral comb electrodes is proposed for high-sensitivity pressure detection in this work. Compared to traditional capacitive pressure sensors with straight plate electrodes, the proposed sensor with the spiral electrodes increases the overlap areas of electrodes sufficiently, the pressure sensitivity can thus be greatly improved. Moreover, the capacitance variation of the proposed sensor is dominated by the change of the overlap area of the electrodes rather than the electrode's distance, the linearity can also thus be improved to higher than 0.99. Theoretical analysis and COMSOL-based finite element simulation have been implemented for principle verification and performance optimization. Simulation results show that the proposed design has a mechanical sensitivity of $1.5 \times 10^{-4}$ m/Pa, capacitive sensitivity of 1.10 aF/Pa, and nonlinear error of 3.63%, respectively, at the pressure range from 0 to 30 kPa. An equivalent experiment has been further carried out for verification. Experimental results also show that both the sensitivity and linearity of capacitive pressure sensors with spiral electrodes are higher than those with straight electrodes. This work not only provides a new avenue for capacitor design, but also can be applied to high-sensitivity pressure detection.

***Keywords:*** *Capacitive pressure sensor; COMSOL simulation; Spiral comb electrodes; Mechanical sensitivity*

**Introduction**

With the continuous improvement of micro-electro-mechanical systems (MEMS) technology, MEMS pressure sensors achieved excellent performance with high integration, high stability, and low cost (Mohankumar et al. 2019; Yu et al. 2015), by combining micromachining methods and integrated circuit technology. They have been widely used in many fields such as automobiles, communications, defense, and biomedical (Jongsung et al. 2016; Wan et al. 2018). Compared with other pressure sensors, capacitive pressure sensors have potent functions of low-temperature drift, quick dynamic response, and low power consumption, which play an essential role in the high-precision pressure sensing field (Bin and Huang 1987; Chiang et al. 2007; Maheshwari and Saraf 2008; Shi-Yu et al. 2018). In addition, these pressure sensors also show the advantages of simple preparation, easy signal acquisition and large dynamic range, so the research on novel capacitive pressure sensors has been the focus of attention in recent years (Ganji and Shahiri-Ta Ba Restani 2013; Jindal et al. 2018; Merdassi et al. 2016; Zhang et al. 2019).

The detection principle of capacitive pressure sensors is realized by changing the distance between the electrodes, while the distance change is nonlinear (Daigle et al. 2007). Moreover, the capacitance of these sensors is small due to the limitation of the electrode plate size, which makes the sensor sensitive to electromagnetic interference and reduces the measurement sensitivity (Nagata et al. 1992). In order to improve the anti-interference function and sensitivity of capacitive pressure sensors, many studies focused on structure design. An innovative design of a capacitive pressure sensor clamp-based was presented, utilizing a $Si_3N_4$ layer as the dielectric medium. The simulation results showed that the sensitivity obtained was greatly improved (Rao et al. 2020). Balavalad et al. established the slotted and perforated microstructure of the capacitive pressure sensor simulation model based on the conventional capacitive sensor respectively (Balavalad 2015). The simulation results showed that the perforated sensor could promote a range of pressure detection. The slotted sensor reduced the influence of the residual stress and achieve better sensitivity. Shivaleela et al. proposed a novel design by adding springs between plates on the MEMS pressure sensor (Shivaleela et al. 2017). The research of adding 1-spring, 4-springs, and 9-springs in the capacitive pressure sensor is designed and simulated, and the 9-spring pressure sensor model obtained the highest sensitivity of $5.12e^{-14}$ f/Pa. However, the above structure improves the sensitivity only in the small deflection range. In addition, some researchers focused on optimizing the sensing diaphragm. Akhil et al. changed the shape of the sensor diaphragm from the ordinary flat one to the central bossed structure (Ramesh and Ramesh 2015). They found that the bossed diaphragm improved the sensitivity, but the range was limited. These designs only improve the sensitivity but not the nonlinearity of the sensor.

Aiming to improve the linearity of capacitive pressure sensors, there have been many studies on the optimization of electrode structure recently, including the optimization of electrode shape (Bin and Huang 1987), adding a reference electrode (Benzel et al. 1994), and adopting multi-electrode structures (Kim et al. 1997). Zhen et al. established the model of a capacitive pressure sensor with a non-coplanar comb electrode structure based on COMSOL (Liu et al. 2019). The capacitance of the sensor was dominated by the change of the overlap area of comb electrodes. Xue et al. proposed a new design of a capacitive pressure sensor composed of comb electrodes and island-diaphragm structure (Li et al. 2019). The overlap length change of the comb electrodes in a particular direction leads to the capacitance variation. The simulation results indicated an increase of linearity. Another innovative structure of capacitive pressure sensors based on comb electrodes was proposed, which reaches high linearity with a wide dynamic range via vertically arranged comb fingers (Ganjalizadeh et al. 2014). Furthermore, Chitra et al. utilized a mechanical coupling to connect the pressure sensing diaphragm to the movable comb plate and separate the pressure sensing diaphragm from the movable comb plate (Chitra and Ramakrishnan 2014). Thus, the sensor has both a sensing element and an actuator, providing an independent diaphragm to sense the pressure and improving sensitivity and linearity.

Above mentioned reports indicate that only few studies investigated the influence of the size of the sensor parameter. Thus, this work focuses on exploring the effect of the main dimension parameters on sensor performance and optimizing the simulation model by analyzing the simulation results. We designed an innovative capacitive pressure sensor that combines a comb structure and diaphragm. The capacitance variation of the proposed sensor is dominated by the change in the overlap area of the spiral electrodes. The spiral electrodes used in capacitive pressure sensors were verified through simple equivalent experiments to significantly improve the sensor sensitivity.

**Materials and Methods**

*Sensor design*

Based on the advantages of the low impedance, high signal-noise ratio, and high sensitivity brought by the design of comb electrode structure (Liu et al. 2019), we designed a capacitive pressure sensor with a novel comb electrode structure. As shown in Fig. 1, considering improving the electrode structure to optimize the sensor performance, the traditional straight plate comb electrodes were further modified into the innovative structure of spiral comb type electrodes.

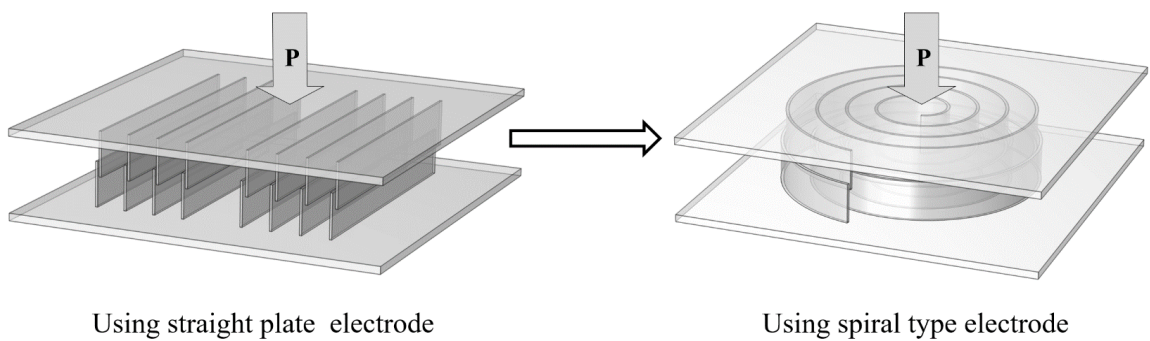

Using straight plate electrode          Using spiral type electrode

**Fig. 1** Straight plate comb electrodes are modified into spiral comb-type electrodes.

The integral structure of the pressure sensor is shown in Fig. 2. The sensor contains four components: diaphragm, mechanical coupling, moving plate, and fixed plate. The moving plate structure includes a sensing diaphragm and moving comb electrodes. The fixed electrodes and sensing diaphragm comprise the fixed plate structure. As the sensing diaphragm is under operating pressure, the overlap areas between the movable comb electrode and the fixed comb electrode are changed by the deflection of the sensing diaphragm, thus determining the capacitance change of the sensor.

In addition, a cylindrical mechanical coupling was designed to connect the diaphragm and moving plate. Compared with applying pressure directly to the sensing diaphragm, this design provides an independent diaphragm to sense pressure. The deflection was transformed into linear displacement through a mechanical coupling element, which improves the sensitivity and accuracy of the sensor (Chitra and Ramakrishnan 2014). The diaphragm size is designed to be 100×100×30 μm, which not only ensures sufficient thickness of itself but also prevents local deformation under pressure. The size of the mechanical coupling is small so that the influence of the coupling on the deformation of the sensing diaphragm is ignored (Bhol 2017).

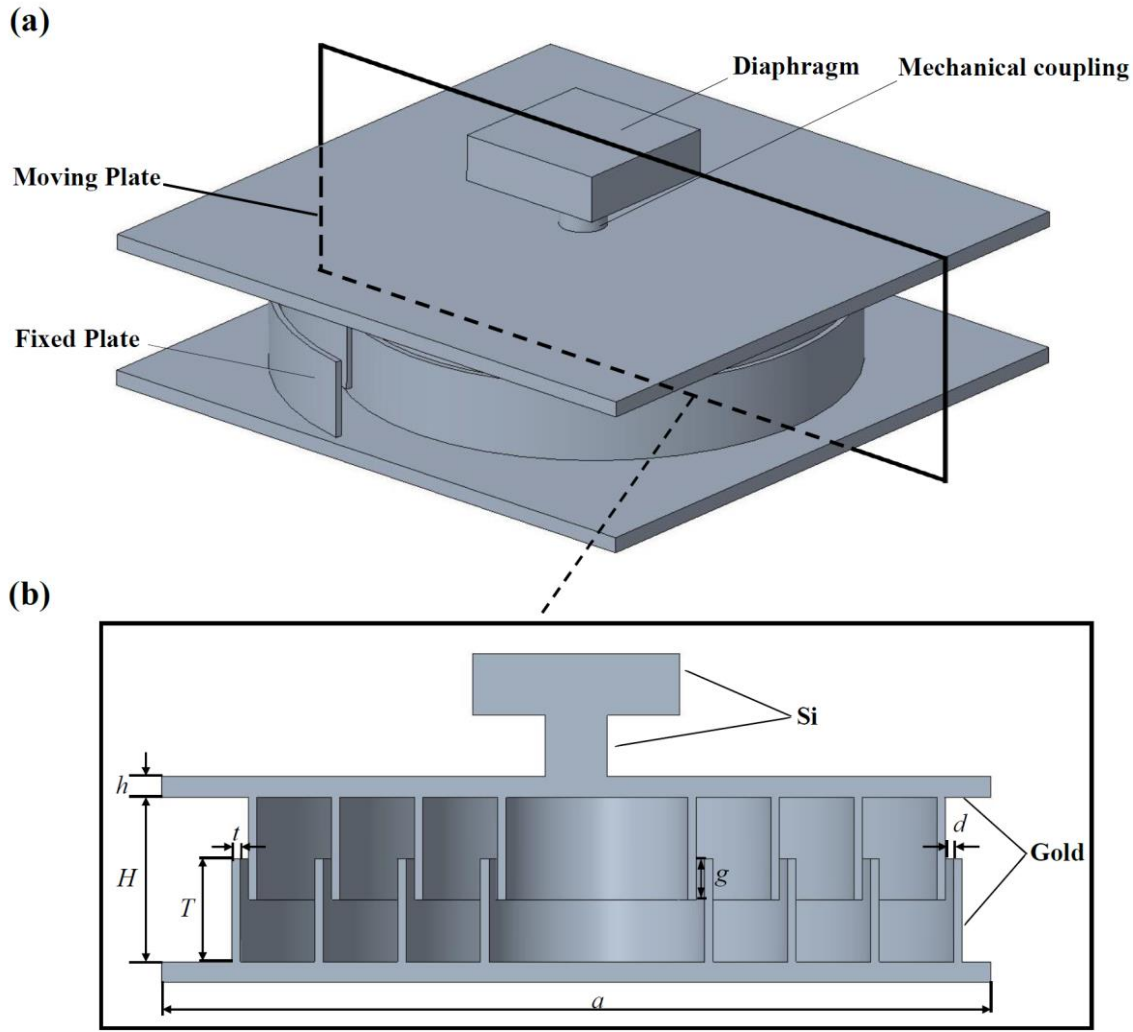

**Fig. 2** The integral design of the proposed sensor: (a) geometry structure diagram of the sensor; (b) cross-section with dimensional parameters of the sensor.

*Theoretical analysis*

The material selected for the sensor has a large Young's modulus and bending stiffness. The deflection of the sensing diaphragm would be far less than its own thickness. Therefore, the small deflection theory of thin plate was chosen for the subsequent analysis and calculation. In this way, the problem was simplified to the pure bending deformation theoretical assumption, thus significantly reducing the complexity of theoretical calculation. Furthermore, the sensing diaphragm of the capacitive pressure sensor is square or round. Compared with the circular diaphragm, the square diaphragm produces minor deviation and higher sensitivity after applying pressure, making it more suitable for detecting slight pressure (Yu et al. 2013). Consequently, the square diaphragm was used as the sensing diaphragm in the proposed sensor.

For a square diaphragm, the small deflection w(x, y) of any point is determined by the following equation (He et al. 2013):

$$w(x, y) = \frac{0.0213 p a^4 (1-\frac{x^2}{a^2})^2 (1-\frac{y^2}{a^2})^2}{D} \tag{1}$$

where $D$ is the flexural rigidity of the sensing diaphragm and is expressed as:

$$D = \frac{Eh^3}{12(1-v^2)} \tag{2}$$

where $E$ represents Young's modulus of the diaphragm material, $h$ denotes the diaphragm thickness, and $v$ is Poisson's ratio of the diaphragm material.

According to Equation (1), the maximum deflection of the diaphragm is in the center of the diaphragm ($x=0$, $y=0$), which is obtained as:

$$w_{max} = w(x, y) = \frac{p a^4}{47 D} \tag{3}$$

The change of the overlap area of the comb electrodes caused by the deflection of the sensing diaphragm under pressure results in the change of capacitance. As a result, the sensor capacitance $C$ can is calculated as:

$$C = \frac{\varepsilon_0 \varepsilon_r l (g+w)}{d} \tag{4}$$

where $\varepsilon_0$ is the dielectric constant, $\varepsilon_r$ is the relative dielectric constant, $l$ denotes the overall overlap length of comb electrodes, $g$ represents the overlap depth of comb electrodes, and $d$ is the gap between adjacent comb tooth electrodes. The above equations indicate that the deflection of the sensing diaphragm is directly proportional to the applied pressure.

*Finite element model*

Considering the working principle of the capacitive pressure sensor, a simulation model was established based on COMSOL software. The Solid Mechanics Module and Electrostatics Module were utilized to simulate the working process of this model, and Electromechanical Forces Multiphysics Field was added to couple these two modules. Considering the actual working condition of the capacitive pressure sensor, a cube was added between two plates and given the air material property to act as a dielectric between the electrodes.

The Solid Mechanics Module was utilized to simulate the sensing diaphragm deflection in the founded model, where the applied pressure was simulated with boundary pressure. The source boundary condition with pressure of 50 kPa was added to the diaphragm. Static constraint was added to the sensor. The sensor capacitance was simulated using the Electrostatics Module. The spiral electrodes were placed in a cuboid air domain to simulate the electricity between the electrodes. The movable electrode was specified at a positive potential of 1 V. The fixed electrode was set as 0 V. The rest of the sensor were set to be insulated. The boundary condition of grounding was applied to the lower surface of the fixed plate.

The solver in COMSOL selects the steady-state to analyze the statics. However, to improve the convergence of the nonlinear model and to facilitate the analysis of simulation results under different pressures, an additional scan in the research extension was added to the steady-state solver. In addition, the applied pressure of 50 kPa was divided into ten times for linear solution. Considering the requirement of thermal stability, electrical conductivity and uniformity of the material, the comb structure is made of gold (Bhol 2017). The material and the initial main dimension parameters used for the sensor are concluded in Tables 1 and 2, respectively.

**Table 1** Material properties of the capacitive pressure sensor used in the model.

| Material | Parameter (unit) | Default value |
|---|---|---|
| Gold | Density (Kg/m$^3$) | 19300 |
| | Young's modulus (GPa) | 70 |
| | Poisson's ratio | 0.44 |
| Silicon | Density (Kg/m$^3$) | 2320 |
| | Young's modulus (GPa) | 160 |
| | Poisson's ratio | 0.22 |
| Air | Dielectric constant | 1 |

**Table 2** Parameters of the geometric model.

| Parameter | Description | Value (μm) |
|---|---|---|
| h | The diaphragm thickness | 10 |
| a | The diaphragm length | 400 |
| t | The electrode thickness | 4 |
| T | The electrode height | 50 |
| d | The gap between electrodes | 4 |
| g | The overlap electrode length | 20 |

**Results and Discussion**

*Initial simulation*

Through the static analysis of the structure, the deformation of the sensor under stress is explored. Figure 3 illustrates the deflection changes of the whole sensor before and after applying pressure (50 kPa). Figure 3(a) shows that the maximum displacement of the sensor is $1.89 \times 10^{-9}$ μm, indicating that the sensor structure would be affected by its gravity. However, the effect on the model is so small that it is neglected. Figure 3(b) shows a maximum deflection of 1.85 μm < 0.3h = 3 μm under 50 kPa, which is consistent with the slight deflection theory assumed. In addition, Fig. 3 reveals that the maximum deflection appears in the center of the square sensing diaphragm, which verifies the deduction of Equation (1).

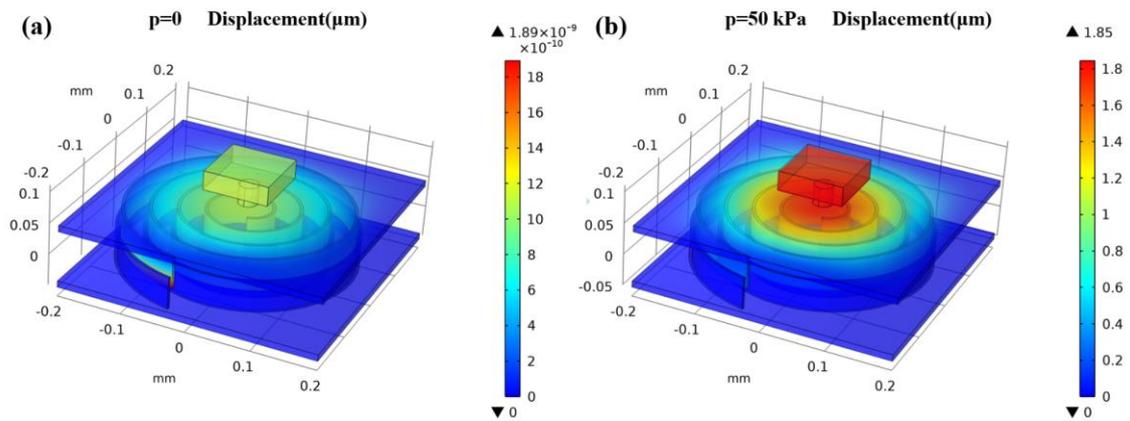

**Fig. 3** Simulation displacement results about the model under applied pressures: (a) 0 kPa and (b) 50 kPa.

The theoretical maximum displacement of the sensing diaphragm under different pressures was calculated by Equation (1). A contrastive analysis with the simulation results was implemented. Figure 4 shows that the simulation value of the maximum deflection is in a positive proportion to the applied pressure, which follows the conclusion obtained from the theoretical calculation above. The mechanical sensitivity of the theoretical value is $3.76 \times 10^{-5}$ μm/Pa, and the mechanical sensitivity of the simulation value is $3.54 \times 10^{-5}$ μm/Pa. The maximum deflection difference between the theoretical and simulation values is 0.11 μm with an error of 5.85%. The results further verify the rationality of the simulation model established by this design.

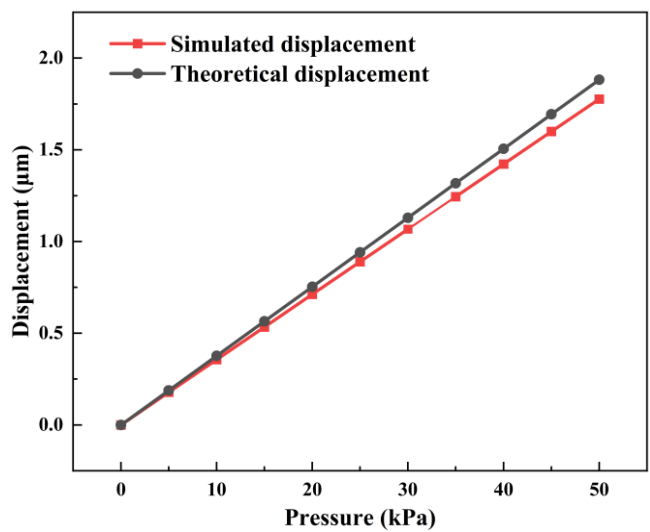

**Fig. 4** Comparison between simulated and theoretical values of the maximum diaphragm deflection for the applied pressure.

Figure 5 exhibits the simulation results of the electric potential distribution and electric field at section Z = 0 under the applied pressure. Figure 5(a) indicates the potential changes due to the deflection deformation of the sensing diaphragm and comb electrode. Spiral electrodes have a higher potential as the terminal, while the cavity is all-around at low potential. Figure 5(b) shows the consistent results: the spiral electrode is distributed with a large electric field on the cross-section, with a maximum value of $4.9 \times 10^4$ V/m, while the cavity electric field distributed around it is weak with the minimum electric field value of 0.

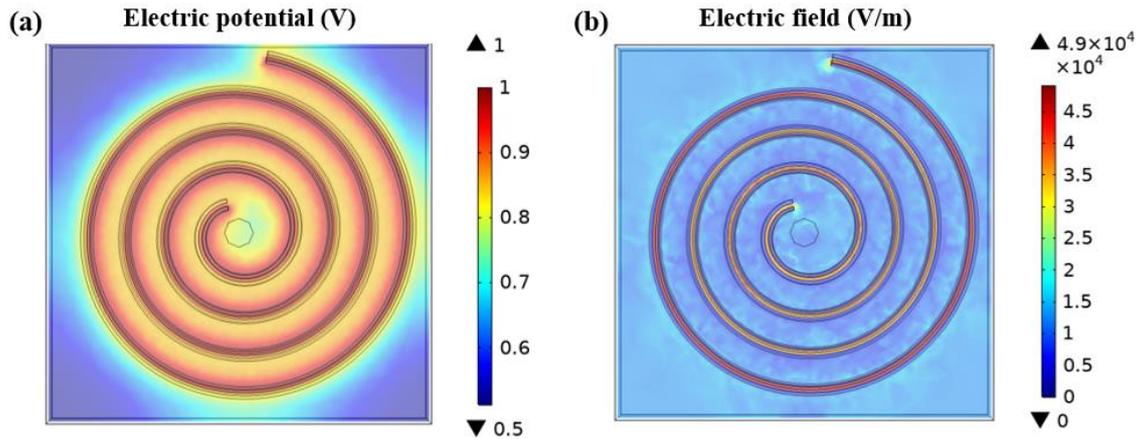

**Fig. 5** Simulation of electric potential distribution and electric field:(a) Potential distribution simulation at the section of Z = 0. (b) Electric field distribution simulation at the section of Z = 0.

Figure 6 shows the theoretical and simulation values of the capacitance changing with the applied pressure. According to the calculation results, the capacitance sensitivity of the theoretical capacitance is 0.10 aF/Pa, while the capacitance sensitivity of the simulation capacitance is also 0.10 aF/Pa. However, in the overall change process, the theoretical capacitance value is 0.001 pF (larger than the simulation capacitance), which is speculated to be caused by the theoretical calculation ignoring the influence of parasitic and edge capacitance on the model. The simulation values nonlinear error is 5.8% with the applied pressure range from 0 to 50 kPa. We speculated that during the deflection of the sensor under pressure, it generated a lateral displacement, resulting in a change in the electrode gap, while theoretical calculation ignored such displacement. Compared with the pressure sensor using the traditional comb electrode, this design significantly improves the relative coverage area of the electrode and achieves higher sensitivity and better linearity due to the innovative structure of the spiral shape.

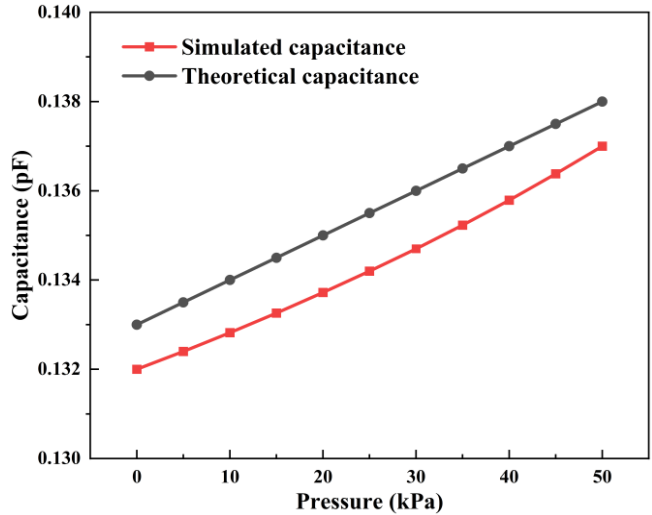

**Fig. 6** Comparison between simulated capacitance and theoretical capacitance for the applied pressure.

*The exploration of the parameter*

Through the analysis of the working principle and structure of the capacitive pressure sensor, the main parameters that affect the performance of the sensor include: sensing diaphragm thickness and length, spiral comb electrodes thickness and height, adjacent electrodes gap, and the relative overlap length. Three single-factor simulation experiments were carried out on each variable to explore the influence law of these six parameters. The rest of the parameters were consistent with the initial simulation values. Each value was selected with the consideration of actual feasibility. The exact test values are shown in Table 3 below. The reason for selecting three values per group to carry out the inquiry experiment is not only to ensure that the three values explore the influence law of the parameters, but also to reduce the number of simulation tests and save the cost of calculation and analysis.

**Table 3** Experimental simulation parameters.

| Parameter | Description | Value (μm) |
|---|---|---|
| $h$ | Sensing diaphragm thickness | 10, 15, 20 |
| $a$ | Sensing diaphragm length | 400, 450, 500 |
| $t$ | Electrode thickness | 3, 4, 5 |
| $T$ | Electrode height | 40, 50, 60 |
| $d$ | The gap between electrodes | 2, 4, 6 |
| $g$ | Overlap electrode length | 10, 20, 30 |

Figure 7 and Figure 8 show the explored simulation results of the six groups of parameters. As shown in Fig. 7 below, the influences of sensing diaphragm thickness h, sensing diaphragm length a, and electrode thickness t on sensor performance were investigated. For the sensing diaphragm thickness h, as the increase of diaphragm thickness, the deflection gradually decreases, as shown in Figs. 7 (a-b). When the deflection thickness size increases to 20 μm, the deflection value is reduced to 0.61 μm, and the mechanical sensitivity is reduced to 1.22 × $10^{-5}$ µm/Pa. In addition, as the thickness of the diaphragm increases, the value of the capacitor increases with the pressure decreases gradually. When the thickness size increases to 20 µm, the capacitive sensitivity is 0.036 aF/Pa. Therefore, in the application of slight deflection, a small thickness of the sensing diaphragm and h =10 µm should be selected as the optimization parameter.

  As for the sensing diaphragm length a and electrode thickness t, the corresponding results were analyzed. Figures 7(c-d) reveal that the center displacement and the capacitance value gradually increase as the sensing diaphragm length increases, which is caused by the increase in the diaphragm area. Therefore, under a reasonable range, a large-diaphragm area should be chosen as the sensing diaphragm. Similarly, Figs. 7(c-d) show that the mean deflection value and capacitance growth decrease with the increase of electrode thickness, thus the small thickness of the electrode size should be selected in order to achieve the result of improving sensor performance. Then, a = 500 µm, t = 3 µm were chosen as the optimized parameters.

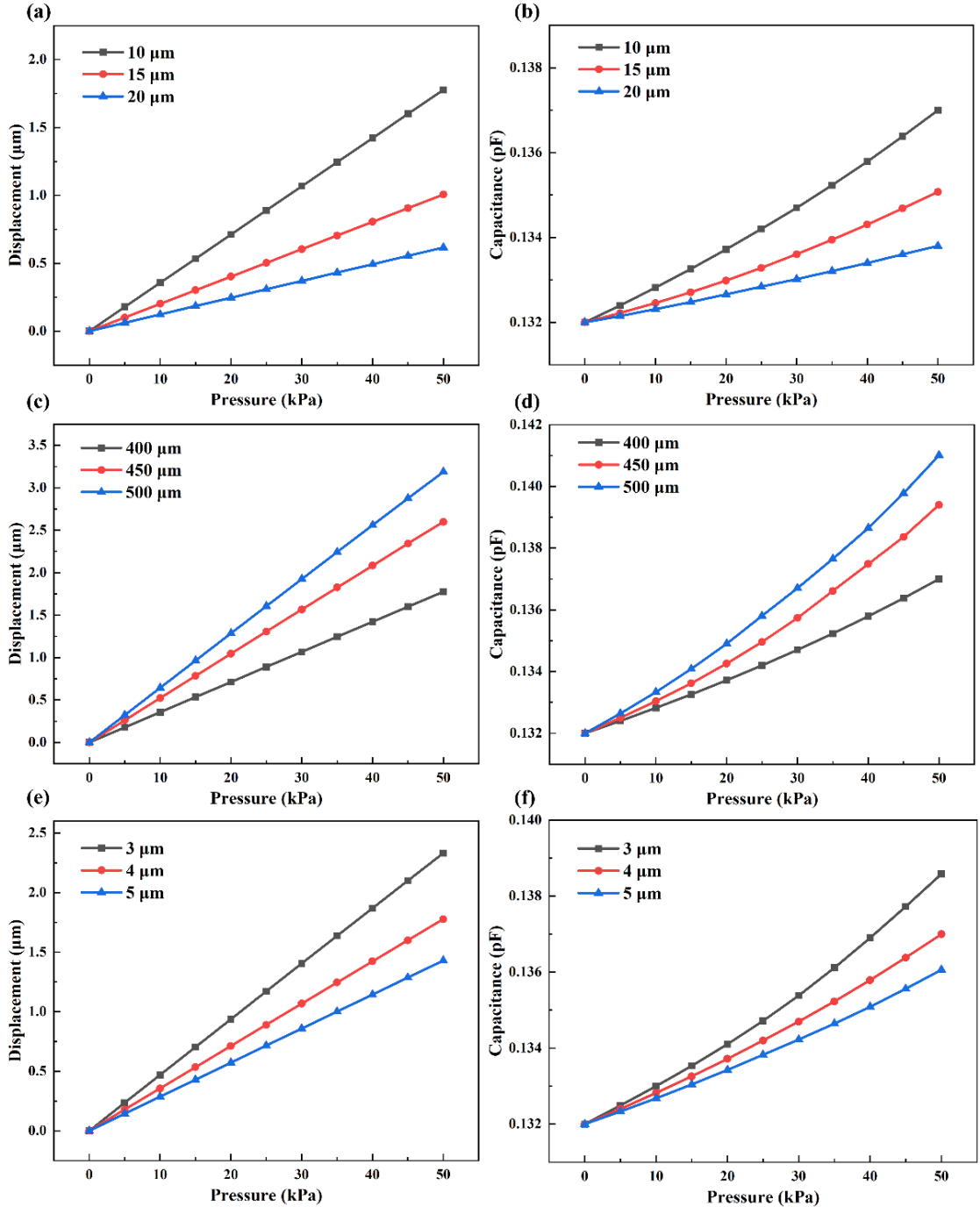

**Fig. 7** Comparison of the center deflection-pressure and capacitance-pressure relationships of (a-b) diaphragm thickness; (c-d) diaphragm length; (e-f) electrode thickness.

Figures 8(a-b) show the influences of electrode height $T$ on sensor performance. The results show that different electrode height has little effect on displacement and capacitance. The corresponding deflection and capacitance value reach the maximum with the increase of electrode height. Therefore, $t = 60$ μm was chosen as the optimization parameter.

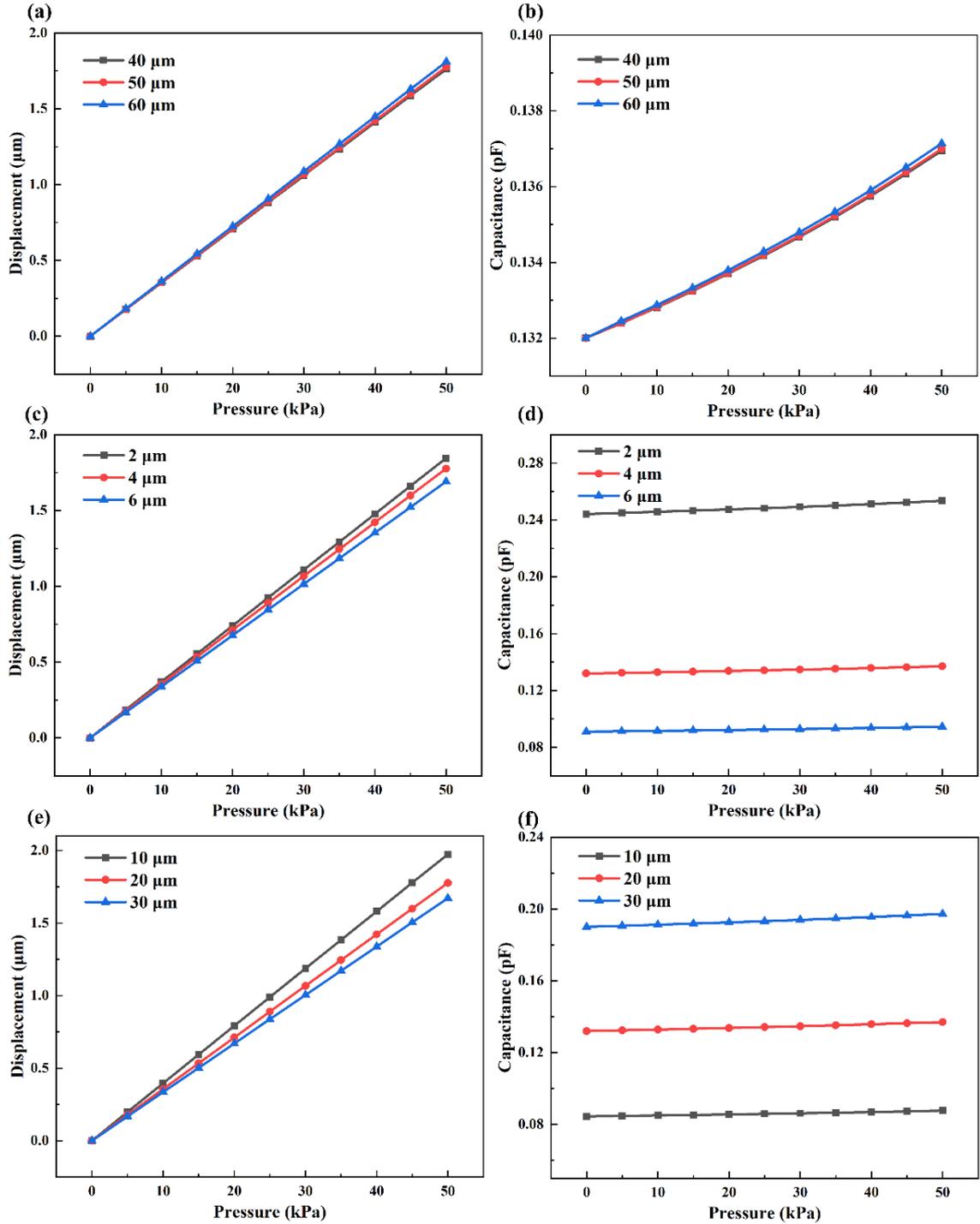

**Fig. 8** Comparison of the center deflection-pressure and capacitance-pressure relationships of (a-b) electrode height; (c-d) gap between electrodes; (e-f) overlap electrode length

The corresponding results of *d* and *g* were analyzed. Figure 8(c) shows that the smaller the comb gap, the greater the deflection value. However, the result difference between different values is small, thus the change of the comb electrode gap has little effect on mechanical sensitivity. Different from the above conclusion, different adjacent electrode gap values significantly influence the capacitance of the sensor. Figure 8(d) shows the initial capacitance value is maximum at 0.244 pF, and the capacitive sensor is maximum at 0.2 aF/Pa with *d* = 2 µm; When $d = 6$ µm, the minimum initial capacitance is 0.09 pF, while the minimum capacitance sensitivity is 0.066aF/Pa. Thus, the smaller the gap between the comb electrodes, the greater the capacitive value and capacitive sensors. Therefore, the value of the electrode gap has an important influence on capacitive sensitivity. Adjacent electrodes should be set to small gaps within a reasonable numerical range to improve sensor performance. Then, $d = 2$ µm was chosen as the optimized parameter.

Figures 8(e-f) show that the deflection value increases with the increase of electrode overlap area electrode thickness. However, the capacitance of the sensor is greatly affected by the different overlap lengths of the electrode. The initial capacitance value is maximum at 0.19 pF, and capacitive sensitivity is maximum at 0.14 aF/Pa at $g = 10$ µm; When $g = 30$ µm, the initial capacitance value is minimal at 0.08 pF, and the capacitive sensor is minimum at 0.06 aF/Pa. Thus, the larger the electrode overlap length value, the greater the initial capacitor value and capacitive sensitivity. The resulting difference between the three sets of data is significant with the effect of the deflection. The overlap length of the electrode directly determines the size of the coverage area, having a significant effect on capacitive sensitivity. Therefore, a more considerable electrode overlap length should be used within a reasonable numerical range to improve sensor performance. So, $g = 30$ µm was chosen as the optimized parameter.

*Optimized results*

The simulation analysis of the optimized model was carried out by replacing the relevant parameters of the sensor model. Figure 9(a) shows that the mechanical sensitivity value of the optimized model is $1.5 \times 10^{-4}$ µm/Pa. Compared with the initial simulation result, the mechanical sensitivity of the sensor using the optimized parameters is improved by about 3.2 times, which greatly increases the sensitivity of the pressure sensor. Figure 9(b) depicts the simulated capacitance value of the optimized model that varies with the applied pressure. The initial capacitance value of the model is 0.385 pF, and the capacitance value increases to 0.418 pF under the applied pressure of 30 kPa. The result shows that the capacitance sensitivity of the optimized model is 1.10 aF/Pa, and the nonlinear error of the simulation value is 3.63%. Compared with the initial simulation model results, the performance of the optimized sensor is greatly improved., including a ten times increase in capacitive sensitivity and a 59.8% increase in linearity.

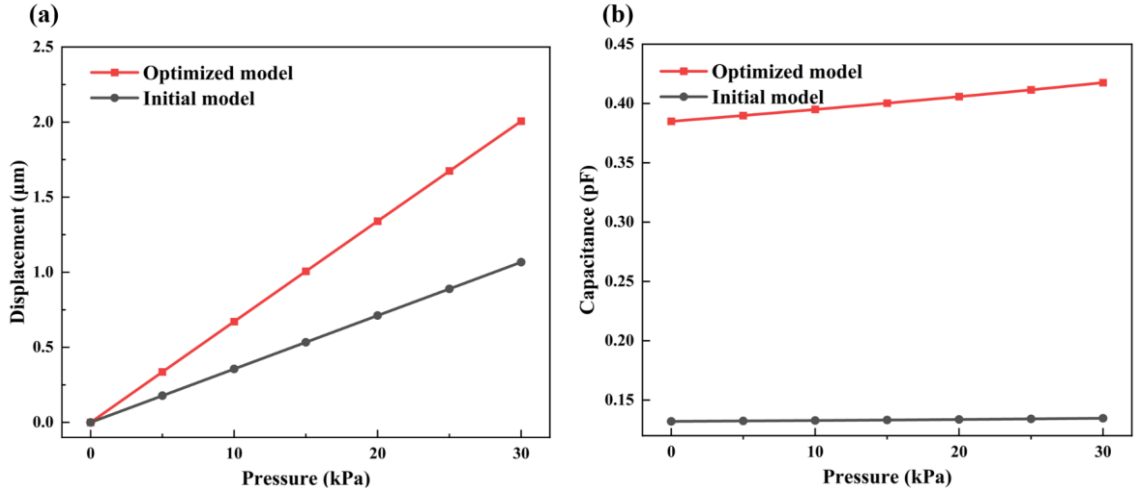

**Fig. 9** Comparison between optimized and initial model: (a) the center deflection-pressure relationship; (b) the capacitance-pressure relationship

To further explore the advantages of this design, we established a model with other parameters (i.e., $h, a, t, T, d, g$) consistent with the model, but using traditional straight combed electrodes. Figure 10 illustrates the comparison between spiral and straight plate electrodes used in sensors. Figure 10 illustrates the comparison between spiral and straight plate electrodes used in sensors. Figure 10(a) indicates that the deflection of the spiral electrode model is slightly larger than the straight electrodes. Figure 10(b) illustrates the simulated capacitance of different electrodes under the applied pressure. The capacitance sensitivity of the traditional model is 0.25 aF/Pa. Compared with the simulation results of the improved model, the capacitance sensitivity of the sensor using the spiral electrodes is increased by 3.4 times, which greatly improves the performance of the capacitive pressure sensor. Moreover, the spiral electrodes used in capacitive pressure sensors greatly increase the overlap area between electrodes so the sensor sensitivity is greatly improved.

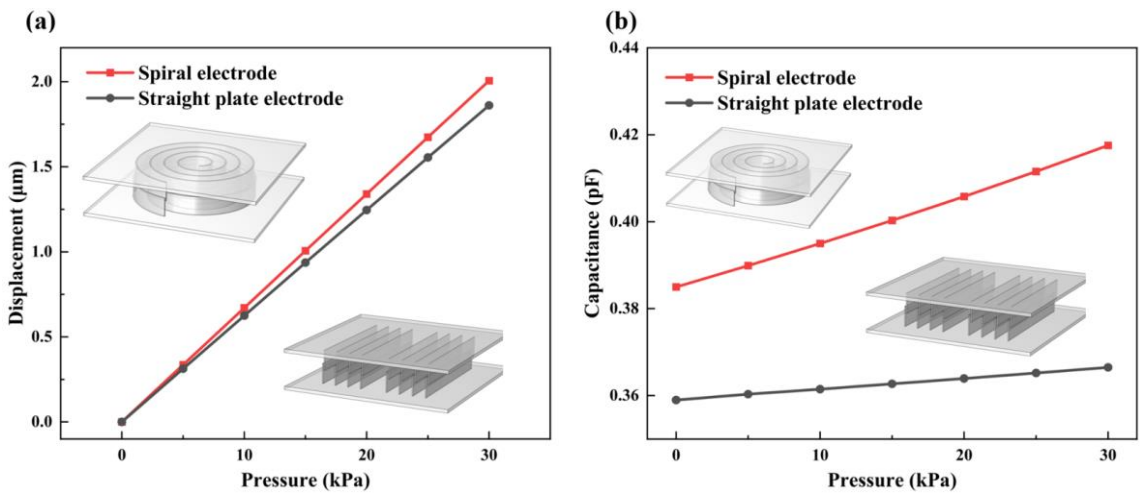

**Fig. 10** Comparison between Spiral and straight plate electrodes: (a) the center deflection-pressure relationship; (b) the capacitance-pressure relationship

**equivalence experiment**

Simulation results show that, capacitive pressure sensors using helical electrodes have higher sensitivity and linearity compared to straight electrode structures. The results are verified by a simple equivalence experiment, which demonstrates that the helical electrode designed in this paper has a higher capacitive sensitivity than the conventional structure when applied to capacitive pressure sensors. Since it is only a simple verification, its own model parameters are amplified to a certain extent in the same direction, and only the difference in the structure is taken as the only variable, so the data appearing in this experiment is only used as a reference for the theoretical verification. The equivalent experimental validation model of the design is illustrated in Figure 11. Figure 11(a) shows the electrode model of the spiral structure, Figure 11(b) shows the electrode model of the comb structure, and Figure 11(c) shows the schematic demonstration of the design of the structure. For the convenience of verification, other basic parameters of the two structures were also set to be consistent. The capacitor housing was prepared using aluminum alloy, and the structure of the mechanical coupling connecting the diaphragm was replaced by a spacer with different heights, which can change the spacing between the upper and lower electrodes to simulate the difference in the overlapping area of the electrodes under the pressure change. The spacers are made of a non-conductive resin to secure and isolate the upper and lower electrodes, ensuring the accuracy of the equivalent experiments.

In the experimental validation using this model, only the sensitivity is considered, and since the simulation of pressure changes is performed by varying the height of the shims, it is only necessary to verify that the electrode model with a spiral structure has a greater change in capacitance value compared to the comb structure model for the same change in overlap thickness. The experimental model and the measurement process are shown in Fig. 12. Fig. 12(a) shows two designed equivalent validation structures, which are fabricated by precision machining and can be removed to install part of the shims with different heights. There are two ways of placing and installing the capacitors, and in the measurement of capacitance data, the disturbing factors can not be eliminated, and the two ways of placing them are set up to measure the capacitance values, the vertical one, which has a small area of contact with the bottom surface, and the horizontal one, which has a large area of contact with the bottom surface. The vertical type, which has a small contact area with the bottom surface, and the horizontal type, which has a large contact area with the bottom surface, are used in the same way for both types of capacitors in the measurement. The process of measuring the capacitance value of a transverse capacitor using a multimeter is illustrated in Figure 12(b), where we measure the value of the capacitance of the horizontal type by intercepting the same length of tin wire, pasting it on the surface of the capacitor, and connecting it to the multimeter through crocodile clips to measure the value.

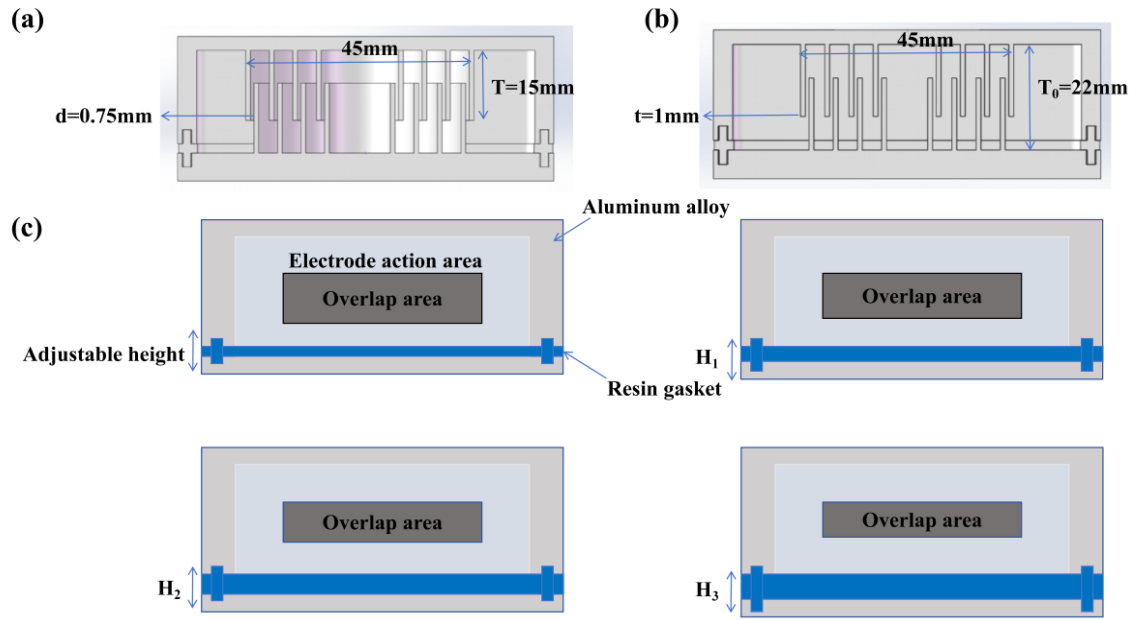

**Fig. 11** Equivalent experimental validation models: (a) electrode model with spiral structure; (b) electrode model with comb structure; $T_0$ is Initial height;(c) schematic diagram of the principle of equivalent structure; $H_3$ is the greater height and $H_1$ is the lesser height, in the diagram $H_3 > H_2 > H_1$

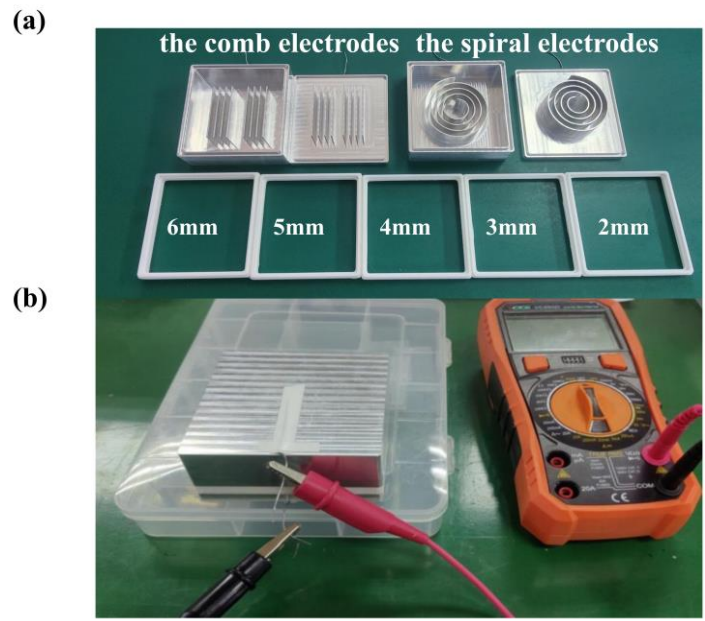

**Fig. 12** Equivalent experimental procedure: (a) experimental device; (b) multimeter to measure the value of capacitance

Figure 13 gives a bar graph comparing the capacitance change with the overlap thickness of the two structures after the experiment, and it can be seen that the change in capacitance value of the spiral-structured electrode is larger than that of the comb structure, both for transverse and vertical measurements.

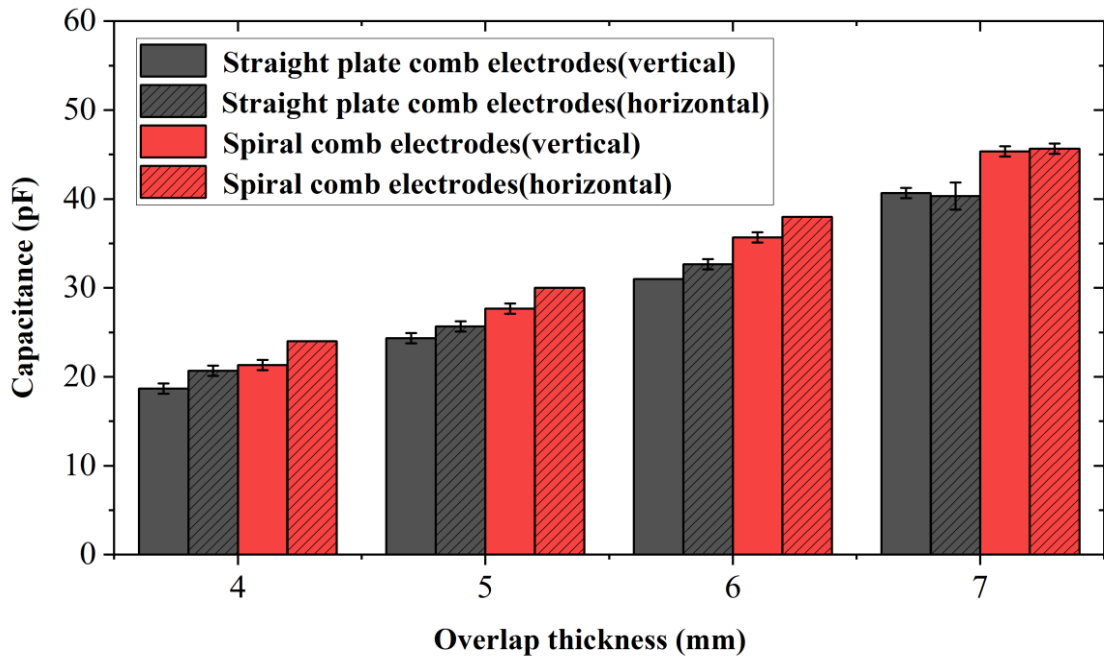

**Fig. 13** Capacitance value versus overlap thickness

Furthermore, a fitting curve was made to characterize the capacitance change versus the overlap thickness as shown in Figure 14(a-b). By solving the fitted straight line, the sensitivities of the two pressure sensor structures under both horizontal and vertical placement methods have been derived from the slope. In the horizontal position, the sensitivity of the spiral comb pressure sensor is 7.20 pF/mm with a linearity $R^2$ of 0.99135, and the sensitivity of the straight comb pressure sensor is 6.53 pF/mm with an $R^2$ of 0.98056. In the vertical position, the sensitivity of the spire comb pressure sensor is 7.43 pF/mm with an $R^2$ of 0.99254, and the sensitivity of the straight comb pressure sensor is 7.19 pF/mm with an $R^2$ of 0.99199. In both measurements, the linearity $R^2$ was calculated and the results showed that the linearity of the spiral structure electrodes was better. Experimental test results show that the sensitivity and linearity of capacitive pressure sensors with spiral electrodes are higher than those with straight electrodes in both measurement modes, which verifies that the performance of capacitive pressure sensors with spiral structure is more advantageous under the same conditions, and confirms the feasibility of electrodes with spiral structure in improving the accuracy.

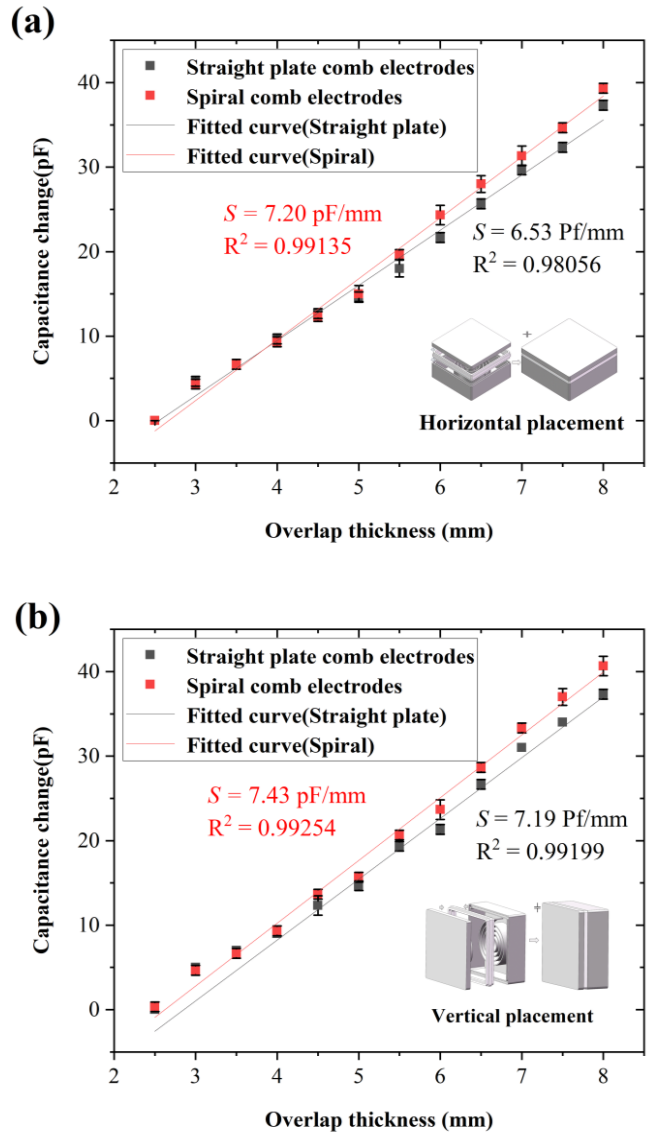

**Fig. 14** The amount of capacitance changes versus overlap thickness: (a) a straight line fitted horizontally; (b) a straight line fitted vertically.

**Conclusions**

This paper proposed an innovative capacitive pressure sensor based on spiral comb electrodes. The reasonability of the established model was identified through theoretical and simulation analysis. The initial size parameters were used to carry out the preliminary simulation of the simulation model. Through a series of simulation explorations, the influence law of the primary dimension parameters of the sensor on the sensor performance was explored. The influence of the six kinds of dimension parameters on the sensor performance was obtained by analyzing the simulation results. The sensor model was optimized according to the analysis results with the improved sensitivity and linearity. The simulated sensor shows a nonlinearity of approximately 3.63% and a sensitivity of 1.10 aF/Pa. Compared with the initial model results, the capacitance sensitivity of the sensor using the spiral electrodes is increased by 3.4 times, which dramatically enhances the performance of the capacitive pressure sensor. Based on the simulation results, the theory is verified by equivalent experiments. Experimental results also show that both the sensitivity and linearity of capacitive pressure sensors with spiral electrodes are higher than those with straight electrodes. This work not only provides a new avenue for capacitor design, but also can be applied to high-sensitivity pressure detection.

**Conflicts of Interest**

The authors declare that there is no conflict of interest regarding the publication of this paper.

**Funding Statement**

This study was supported by the National Key Research and Development Program of China (No. 2021YFB3201200, 2021YFB3201204). Part of this work was also supported by Sichuan Science and Technology Program (No. 2023YFG0041).